\documentclass[aps,preprint,floatfix,superscriptaddress,amssymb]{revtex4-1}

\usepackage[english]{babel}
\usepackage[utf8x]{inputenc}
\usepackage{amsmath}
\usepackage{graphicx}% Include figure files
\usepackage{dcolumn}% Align table columns on decimal point
\usepackage{bm}% bold math
\usepackage[caption=false]{subfig}

\begin{document}
\title{The microscopic origin of DMI in magnetic bilayers and prediction of giant DMI in new bilayers}
\author{Priyamvada Jadaun}
\email{priyamvada@utexas.edu}
\author{Leonard F Register}
\author{Sanjay K Banerjee}
\affiliation{Department of Electrical and Computer Engineering, The University of Texas at Austin, Austin, Texas. USA.}
\date{\today}

\begin{abstract}
Skyrmions are widely regarded as promising candidates for emergent spintronic devices. Dzyaloshinskii-Moriya interaction (DMI) is often critical to the generation and manipulation of skyrmions. However, there is a fundamental lack of understanding of the origin of DMI or the mechanism by which DMI generates skyrmions in magnetic bilayers. Very little is known of the material parameters that determine the value of DMI. This knowledge is vital for rational design of skyrmion materials and further development of skyrmion technology. To address this important problem, we investigate DMI in magnetic bilayers using first-principles. We present a new theoretical model that explains the microscopic origin of DMI in magnetic bilayers. We demonstrate that DMI depends on two parameters, interfacial hybridization and orbital contributions of the heavy metal. Using these parameters, we explain the trend of DMI observed. We also report four new materials systems with giant DMI and new designs for magnetic multilayers that are expected to outperform the best materials known so far. Our results present a notably new understanding of DMI, uncover highly promising materials and put forth novel pathways for the controlled generation of skyrmions.
\end{abstract}


\maketitle

%-----------INTRODUCTION------------
\section{Introduction} As our society's need to store data continuously increases, we will soon reach the performance limits of current memory devices\cite{Luchaire_Fert}. Magnetic skyrmions are widely believed to be one of the most promising candidates for next-generation memory technology. A magnetic skyrmion is a local whirl of spins in a magnetic material with a fixed chirality\cite{Fert_Cros,Fert_Sampaio}. It is topologically protected against deformation into other magnetic states, and against disorder and perturbation. The stability of skyrmions, their small size and their responsiveness to tiny electrical current densities ($\sim10^6 A/m^2$), make these skyrmions ideal for enabling ultra-dense, low-energy memory devices\cite{Fert_Sampaio}. One of the important mechanisms controlling the generation and stability of skyrmions and other magnetic topological defects\cite{Bode_Wiesendanger} is the Dzyaloshinskii-Moriya interaction (DMI)\cite{Dzyaloshinsky, Moriya}.

Despite its fundamental importance, the microscopic origin of DMI in magnetic bilayers is not well understood. Originally derived for magnetic insulators, the theory developed by Moriya\cite{Moriya} does not help us understand the trend of DMI observed in magnetic bilayers. Recent works have tried to uncover the microscopic origin of DMI in magnetic bilayers. One model proposes that the source of DMI in bilayers is the proximity-induced magnetic moments in the heavy metal(HM) layers\cite{Ryu}. However, this proposal is contradicted by another report\cite{Yang_Chshiev_2}. Other interesting studies point out that HM-FM hybridization is a primary factor in controlling the strength of DMI\cite{Belabbes_Manchon,Kashid_Salunke}, with which our model agrees. However, the proposition of \cite{Belabbes_Manchon} that the sign of DMI follows FM 3d band filling is contradicted by the large, negative DMI seen in Ir/Fe\cite{Yang_Chshiev_1, Heinze} and both positive and negative signs of DMI reported for HM/Co\cite{Tolley_Fullerton,Yang_Chshiev_2}. Similarly, the proposal that DMI is controlled solely by the spin-polarized HM 5d states in Pt/FM\cite{Kashid_Salunke} is contradicted by experimental reports of opposite signs of DMI for Pt/Co\cite{Interfacial_Belmeguenai} and Pt/Ni\cite{chen2013tailoring}.

There is thus a gap in our understanding of the relationship of DMI with electronic structure, and of the mechanism by which DMI produces magnetic textures. This gap hampers our ability to design materials for enhanced DMI, or to control the strength or sign of DMI which can help the development of controllable skyrmion technology\cite{Jiang_Hoffmann}. In this paper, we address this gap and present a model to explain the microscopic origin of DMI in metallic bilayers. We calculate the DMI values in a series of heavy metal/ferromagnet (HM/FM) bilayers and explain the trend of DMI observed. In particular, we show that the value of DMI depends on two factors, namely, HM-FM hybridization and HM spin mixing terms. The latter are determined by the contribution of specific HM orbitals to the HM/FM band structure. We emphasize that the HM/FM band structure, and consequently the DMI sign, depends on the choice of both the FM as well as the HM. We also derive a theory for the mechanism by which spin orbit coupling (SOC) generates spin textures. Our results present a significantly new understanding of DMI. They also unveil important avenues to control the strength and sign of DMI, thereby advancing the controlled generation or annihilation of skyrmions.

Real world applications in emergent memory technology would ideally require skyrmions of size $\leq$ 10 nm that are stable at room temperature\cite{Fert_Cros}. Ultrathin HM/FM films have been shown to host small skyrmions ($\sim$ 3-8 nm) stabilized by DMI, but so far they exist only at low temperatures ($<$30 K)\cite{Heinze,Romming}. On the other hand, HM/FM multilayer stacks can host skyrmions at room temperature, but so far they have been larger than 30 nm in diameter\cite{Luchaire_Fert,Soumyanarayanan_Panagopoulos,Woo_Beach,Boulle_Gaudin,Geometrically_Ho}. Recently, it has been proposed that frustrated magnets can also host skyrmions that are stabilized by competing ferromagnetic nearest-neighbor and antiferromagnetic next-nearest-neighbor interactions \cite{frust_th1,frust_th2,frust_th3,frust_th4}. The existence of such skyrmions has been very recently shown for $Gd_2PdSi_3$\cite{frust_ex1}. While these skyrmions are very small in size ( $\sim$ a few nms), so far, they have only been observed at low temperatures ($<$20 K) as well.

In HM/FM bilayers, the size and stability of DMI-stabilized-skyrmions is decided by the interplay between multiple variables, including, the DMI, the exchange constant, the out-of plane anisotropy and dipolar fields\cite{Fert_Cros}. Therefore, to realize skyrmion based emergent memories, it will be important to identify materials with the appropriate combinations of these parameters. As a general rule, a large ratio of DMI to exchange constant encourages a quicker rotation of the spin, and reduces the size of the skyrmion (in the absence of interactions like edge effects)\cite{Fert_Sampaio}. Thus, materials with enhanced values of DMI can host skyrmions that are stable at higher temperatures and small in size. Here, we report new bilayers that demonstrate giant DMI, up to twice the largest value known so far. We also present new material designs for magnetic multilayers that show enhanced DMI. We note that starting with an HM-FM bilayer with a giant interfacial DMI, fine tuning of DMI and exchange values is possible by further engineering of the stacks\cite{Engineering_Dupe, Thickness_Cho}. Large DMI also enables the formation of other chiral spin structures, like chiral domain walls, which are relevant to next-generation data storage devices\cite{Soumyanarayanan}. Our results thus significantly advance the field by presenting new materials which show promise for enabling skyrmion device technology.

%-----------RESULTS ------------
\section{Results} 
\begin{figure}
    \centering
    \includegraphics[width=\columnwidth]{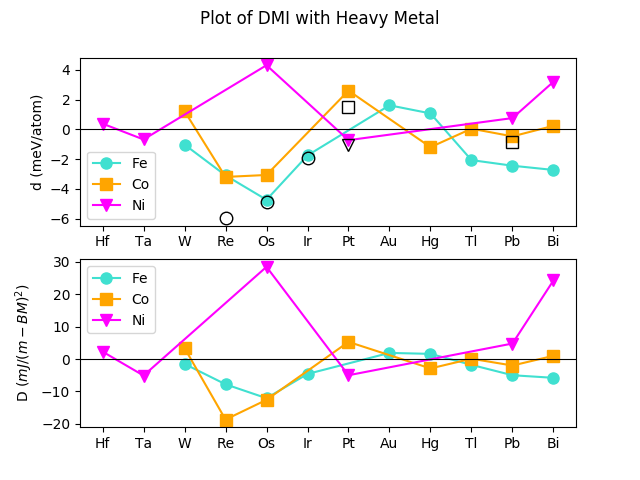}
    \caption{Plots of Dzyaloshinskii-Moriya interaction (DMI) for a series of HM/FM bilayers with HM ranging from Hf to Bi. Our DMI results are plotted with filled markers, whereas corresponding DMI values in literature are plotted using unfilled black markers. Circles represent Fe, squares represent Co and triangles represent Ni. Reference values are taken from \cite{Simon_Szunyogh} for Re/Fe, Os/Fe Ir/Fe; from \cite{Yang_Chshiev_2} for Pt/Co; from \cite{Yang_Chshiev_1} for Pb/Co; and from \cite{chen2013tailoring} for Pt/Ni. The unit of \textit{D} is $mJ/(m-BM)^2$, where $\textit{BM}$ is the Bohr magneton.}
    \label{fig:dmi_val}
\end{figure}

\textit{New materials with giant DMI}- Our results for DMI are shown in Fig.\ref{fig:dmi_val}. Our values show excellent agreement with literature, except for Re/Fe, where our DMI is roughly half that predicted by Simon et al.\cite{Simon_Szunyogh}. A significant result is the giant DMI seen in six bilayers, namely, Re/Fe, Os/Fe, Re/Co, Os/Co, Os/Ni and hexagonal Bismuth(hBi)/Ni. These materials show a DMI (\textit{d}) up to twice the largest currently known values of 1.5 meV/atom for Co/Pt\cite{Yang_Chshiev_2} and -1.9 meV/atom for Ir/Fe\cite{Yang_Chshiev_2}. Predictions of large DMI in Re/Co, Os/Co, Os/Ni and hBi/Ni have not been reported before. As a possible quantum spin Hall material\cite{Reis_Claessen}, hexagonal Bi is particularly noteworthy, as it can help drive skyrmions via a large spin Hall current. A useful materials design strategy for enhancing DMI has been the addition of DMI of opposite sign from successive interfaces\cite{Dupe_Heinze,Hrabec_McGrouther}, as implemented in Pt/Co/Fe/Ir multilayers. Our results make it possible to achieve even larger additive DMIs in simpler multilayer structures, such as Pt/Co/Os and Pt/Co/Re. 

\section{Discussion}
%-----------ORIGIN OF DMI ------------
\textit{Origin of DMI}- We now present the analysis of our results and develop a model for the microscopic origin of DMI. We demonstrate that DMI depends on two parameters, HM-FM hybridization and the presence of specific HM orbitals in the band structure. We find that the strength of DMI is controlled by the relative alignment of HM-FM bands.  Here we show that the sign of DMI is controlled by orbital contributions from the HM rather than the 3d orbital filling of the FM. Our work thus notably alters the current understanding of DMI in magnetic bilayers.

\begin{figure}
    \centering
    \includegraphics[width=\columnwidth]{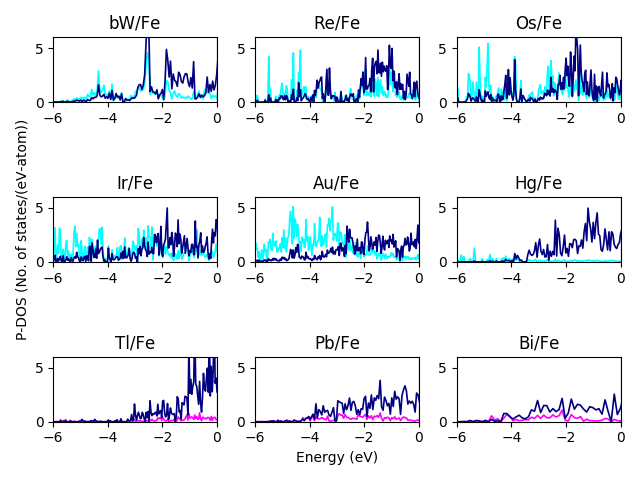}
    \caption{Plot of projected density of states (p-DOS) and band alignment for HM/Fe bilayers. P-DOS for Fe-3d is plotted in navy blue while p-DOS for HM-5d is plotted in cyan and that for HM-6p is plotted in magenta. Fermi energy is set at 0 eV.}
    \label{fig:pdos_HM/Fe}
\end{figure}

\begin{figure}
    \includegraphics[width=\columnwidth]{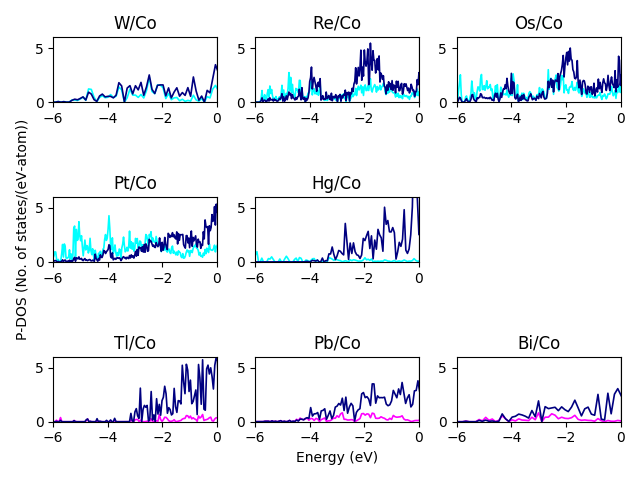}
    \caption{Plot of projected density of states (p-DOS) and band alignment for HM/Co bilayers. P-DOS for Co-3d is plotted in navy blue while p-DOS for HM-5d is plotted in cyan and that for HM-6p is plotted in magenta. Fermi energy is set at 0 eV.}
    \label{fig:pdos_HM/Co}
\end{figure}

%-----------DMI STRENGTH------------
\textit{Origin of DMI strength}- We now show that the strength of DMI (\textit{d}) is principally determined by the HM-FM hybridization. To demonstrate this, we examine the projected density of states (p-DOS), plotted in Fig.\ref{fig:pdos_HM/Fe} and Fig.\ref{fig:pdos_HM/Co}, and relative band alignments between the HM and FM layers. We define band overlap (BO) as the amount of overlap exhibited by the FM 3d bands with the HM 5d bands (6p bands) for the 5d (6p) series of HMs. We expect that the greater the BO, the greater the HM-FM hybridization. Upon visual inspection, we clearly see that \textit{d} strongly depends on BO. In particular, a large BO is a necessary but not sufficient condition to obtaining large \textit{d}. Since Fe, Co and Ni have d6-d8 electron configurations, the 3d-5d band overlap is maximized for Re, Os, Ir and Pt (d5-d8), giving rise to maximum DMI strength for these materials. Similarly for HM-6p elements, \textit{d} increases for increasing BO. There are notable outliers, namely, Ir/Fe, Pt/Co and Pt/Ni (and hBi/Co for HM-6p) where the DMI obtained is smaller than what would be expected from BO alone. We explain the reason for these exceptions later.

For the HM-5d/Fe series, we observe a trend of $|d_{\beta W/Fe}|<|d_{Au/Fe}|<|d_{Re/Fe}|<|d_{Os/Fe}|$. This is easily explained as the BO also increases in this order, as shown in Fig.\ref{fig:pdos_HM/Fe}. Similarly, for the HM-5d/Co series, we note a trend of $|d_{Hg/Co}|\sim|d_{\alpha W/Co}| < |d_{Pt/Co}| < |d_{Os/Co}|\sim|d_{Re/Co}|$. This is explained by the large BO seen for Pt/Co, Os/Co and Re/Co, as shown in Fig.\ref{fig:pdos_HM/Co}. The HM-Ni series also follows similar trends with increasing \textit{d} for increasing BO. As noted before, Ir/Fe, Pt/Co and Pt/Ni have a smaller \textit{d} than one would expect simply from BO. The only other outlier is Hg, which both for Fe and Co has a DMI which is small, but larger than expected from simply 3d-5d BO. Hg/FM results suggest that there might be other, smaller contributions to HM/FM hybridization in addition to the 3d-5d BO. For the HM-6p/FM series, the observed trend of $|d_{Tl/Fe}|\lesssim|d_{Pb/Fe}|\lesssim|d_{Bi/Fe}|$ and $|d_{Tl/Co}|<|d_{Bi/Co}|<|d_{Pb/Co}|$ both correlate well with the increasing order of BO. Overall, our results show that HM-FM band overlap has a significant control over the strength of DMI. We explain the cause of this relationship later.

%-----------DMI SIGN ------------ 
\textit{Origin of DMI sign}- The sign of DMI is important as it decides the sense of rotation of a magnetic texture. We note that DMI's sign does not follow FM 3d band filling and can in fact change even for the same FM. Here we investigate the origin of this sign. We derive a theory for the mechanism by which SOC generates magnetic textures, and demonstrate that the sign of DMI is controlled by the relative presence of HM orbitals in the bandstructure. Our findings bring to light the source of the DMI's sign and point to new pathways for controlling skyrmion chirality, with significant impact to future technology.

\begin{figure}
    \centering
    \includegraphics[width=0.4\columnwidth]{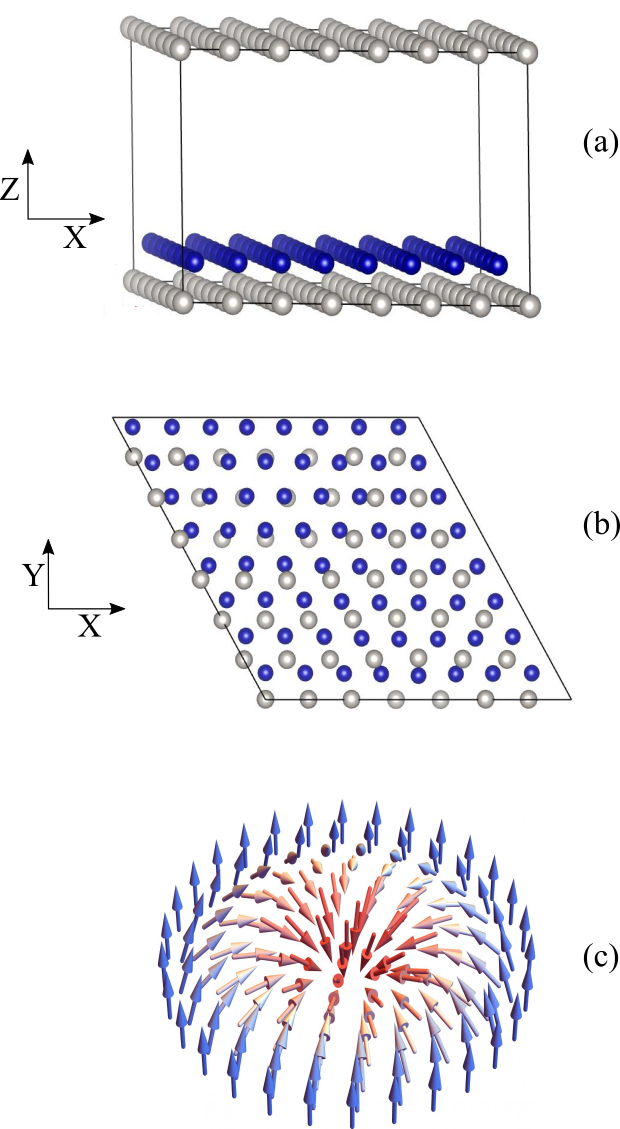}
    \caption{(a,b) Schematic of our structure of Pt/Co bilayer. (a) The front view of the pre-relaxed structure, which includes 10 \AA of vacuum. (b) Top view of the post-relaxed structure. (c) Schematic of a N$\acute{e}$el skyrmion }
    \label{fig:str}
\end{figure}

We use first order perturbation theory to derive the mechanism by which SOC leads to the creation of magnetic textures. We build upon an important previous work on the Berry phase theory of DMI\cite{Freimuth,Hanke_Mokrousov}. According to this Berry phase theory, the rotation of the magnetic moment of an electron alters the free energy, via exchange interaction, and gives rise to DMI. Here we demonstrate how SOC leads to the rotation of magnetic moment in the first place, which subsequently creates magnetic textures and generates DMI. Our derivation applies specifically to the hedgehog-like (N$\acute{e}$el) skyrmions found in magnetic multilayers\cite{Hoffman_Bader}, as shown in Fig.\ref{fig:str}.

%-----------DERIVATION ------------
\textit{Theory for magnetic textures}-Consider a HM/FM bilayer system. In the absence of SOC, the FM has a constant magnetization. Let $\psi^0_{\vec{k}n}$ represent the unperturbed wavefunction of an electron delocalized over this bilayer. 
\begin{equation}\label{eq1}
\hat{H}_0 |\psi^0_{\vec{k}n} \rangle = E^0_{\vec{k}n} |\psi^0_{\vec{k}n} \rangle
\end{equation}
We now turn on SOC, such that $\hat{H} = \hat{H}_0 + \hat{H}_{SOC}$. The new eigenstate representing the electron hopping between HM and FM is given by
\begin{equation}\label{eq2}
|\psi_{\vec{k}n} \rangle = |\psi^0_{\vec{k}n} \rangle + \sum_{\vec{q}, m} \frac{\langle \psi^0_{\vec{k}+\vec{q}m} | \hat{H}_{SOC} | \psi^0_{\vec{k}n} \rangle }{E^0_{\vec{k}n} - E^0_{\vec{k}+\vec{q}m}} |\psi^0_{\vec{k}+\vec{q}m} \rangle 
\end{equation}
Without loss of generality, we can assume the unperturbed FM has a magnetization along the z axis and the perturbed magnetic moment rotates in the XZ plane. We now consider the component of magnetic moment of the perturbed state along $\vec{x}$, up to first order
\begin{equation}\label{eq3}
    \langle \hat{s}_x \rangle =  \sum_{\vec{q}, m}   \langle \psi^0_{\vec{k}n} |\hat{s}_x |\psi^0_{\vec{k}+\vec{q}m} \rangle \frac{\langle \psi^0_{\vec{k}+\vec{q}m} | \hat{H}_{SOC} | \psi^0_{\vec{k}n} \rangle }{E^0_{\vec{k}n} - E^0_{\vec{k}+\vec{q}m}} + C. C.
\end{equation}
Taking the long wavelength limit of $\vec{q} \rightarrow 0$.
\begin{equation}\label{eq4}
    \langle \hat{s}_x \rangle =  \sum_{m \neq n}   \langle \psi^0_{\vec{k}n} |\hat{s}_x |\psi^0_{\vec{k}m} \rangle \frac{\langle \psi^0_{\vec{k}m} | \hat{H}_{SOC} | \psi^0_{\vec{k}n} \rangle }{E^0_{\vec{k}n} - E^0_{\vec{k}m}} + C. C.
\end{equation}
Writing $\hat{s}_x$ and $\hat{H}_{SOC}$ in the spin basis ($|\uparrow\rangle, |\downarrow\rangle$):
\begin{equation}\label{eq6}
    |\uparrow\rangle = 
    \begin{bmatrix}
    1   \\
    0
    \end{bmatrix}
    |\downarrow\rangle = 
    \begin{bmatrix}
    0   \\
    1
    \end{bmatrix}
    \hat{s}_x =
    \begin{bmatrix}
    0 & 1 \\
    1 & 0
    \end{bmatrix}
    \hat{H}_{SOC} = 
    \begin{bmatrix}
    \lambda \hat{l}_0/2 & \lambda \hat{l}_-/2 \\
    \lambda \hat{l}_+/2 & -\lambda \hat{l}_0/2
    \end{bmatrix}
\end{equation}
where $\hat{l}_\pm \equiv \hat{l}_x \pm \iota \hat{l}_y \quad \hat{l}_0 \equiv \hat{l}_z$. We can write $\hat{l}_\pm$ in the d orbital basis ($d_{z^2}, d_{xz}, d_{yz}, d_{xy}, d_{x^2-y^2}$) as,

\begin{equation}\label{eq7}
\hat{l}_+ = 
\begin{bmatrix}
0   &\sqrt{3}    &\iota\sqrt{3} &0  &0 \\
-\sqrt{3}   &0  &0  &\iota  &1 \\
-\iota\sqrt{3}  &0  &0  &1  &-\iota \\
0   &-\iota &-1  &0  &0 \\
0   &-1 &\iota  &0  &0 \\
\end{bmatrix}
\hat{l}_- =
\begin{bmatrix}
0   &-\sqrt{3}    &\iota\sqrt{3} &0  &0 \\
\sqrt{3}   &0  &0  &\iota  &-1 \\
-\iota\sqrt{3}  &0  &0  &-1  &-\iota \\
0   &-\iota &1  &0  &0 \\
0   &1 &\iota  &0  &0 \\
\end{bmatrix}
\end{equation}
For an unperturbed wavefunction $\psi^0_{\vec{k}n}$ in the spin up state, the only nonzero contributions to $\langle \hat{s}_x \rangle$ come from $\psi^0_{\vec{k}m}$ in the spin down state.
\begin{equation}\label{eq8}
\begin{split}
|\psi^0_{\vec{k}n} \rangle = |\phi^0_{\vec{k}n},\uparrow \rangle;  |\psi^0_{\vec{k}m} \rangle = |\phi^0_{\vec{k}m},\downarrow \rangle \\
\langle \hat{s}_x \rangle = \lambda/2\sum_{m \neq n}  \frac{\langle \phi^0_{\vec{k}m} | \hat{l_+} | \phi^0_{\vec{k}n} \rangle }{E^0_{\vec{k}n} - E^0_{\vec{k}m}} + C. C.
\end{split}
\end{equation}
where $\phi^0_{\vec{k}}$ is the orbital part of the wavefunction.
Similarly, for an unperturbed wavefunction $\psi^0_{\vec{k}n}$ in the spin down state, we get:
\begin{equation}\label{eq9}
\begin{split}
|\psi^0_{\vec{k}n} \rangle = |\phi^0_{\vec{k}n},\downarrow \rangle;  |\psi^0_{\vec{k}m} \rangle = |\phi^0_{\vec{k}m},\uparrow \rangle \\
\langle \hat{s}_x \rangle = \lambda/2\sum_{m \neq n}  \frac{\langle \phi^0_{\vec{k}m} | \hat{l_-} | \phi^0_{\vec{k}n} \rangle }{E^0_{\vec{k}n} - E^0_{\vec{k}m}} + C. C.
\end{split}
\end{equation}
Rotation of the magnetic moment of an electron, delocalized over the HM/FM interface, is caused by the spin orbit terms, $\hat{l}_+\cdot\hat{s}_-$ and $\hat{l}_-\cdot\hat{s}_+$. To obtain a continuous, self-sustained rotation in real space, spin up and spin down states should be rotated such that they pick up magnetic moments in opposite directions along the x axis. This condition requires $\langle \phi^0_{\vec{k}m} | \hat{l_+} | \phi^0_{\vec{k}n} \rangle$ to be opposite in sign to $\langle \phi^0_{\vec{k}m} | \hat{l_-} | \phi^0_{\vec{k}n} \rangle$. Inspecting Eq.\ref{eq7}, we conclude that there are only three such terms, namely, $\langle d_{xz}| \hat{l}_\pm | d_{z^2} \rangle$, $\langle d_{xy}| \hat{l}_\pm | d_{yz} \rangle$ and $\langle d_{x2-y2}| \hat{l}_\pm |d_{xz} \rangle$.

To further illustrate this mechanism, we consider a wavefunction hopping across the HM/FM interface. In the presence of SOC, whenever the electron hops to the HM atom, the three SOC spin mixing terms cause d-orbital transitions and rotate its magnetic moment (creating a swirling spin texture). As this electron hops to the FM, its rotated magnetic moment perturbs the wavefunctions of the FM via exchange interaction, altering the free energy and generating DMI\cite{Freimuth}. The stronger the HM-FM hybridization, the stronger the perturbation of energy due to the hopping electron, the larger the DMI. The sign of DMI is decided by the interplay between these three SOC spin mixing terms and the strength of DMI is decided by the HM-FM hybridization. We note that this result does not support the model of \cite{Kashid_Salunke} where the source of DMI is identified to be transitions between $d_{xz}$ and $d_{yz}$ orbitals.

\textit{Orbital control of DMI sign}- We now employ our theory to explain the trends of DMI sign observed in our calculations, thereby shining light on previously unexplained materials behaviour. According to our theory, three SOC transition terms, generate three separate DMI terms. The direction of each SOC transition will decide the sign of the corresponding DMI term, with opposite directions of transition leading to opposite DMI signs. The direction and strength of a transition, in turn, depends on the relative presence of HM orbitals ($d_1$ and $d_2$) in the band structure, before SOC is turned on. Specifically, the stronger the presence of $d_1$, and the weaker the presence of $d_2$, the stronger the transition from $d_1$ to $d_2$ would be, once SOC is turned on. The result of these inferences is that if we compare materials with similar crystal structures, their band structures and energy denominators (${E^0_{\vec{k}n} - E^0_{\vec{k}m}}$) are similar too. Consequently, it is the variation of relative presence of HM d orbitals that controls the net DMI sign. 

\begin{figure}
    \centering
    \includegraphics[width=\columnwidth]{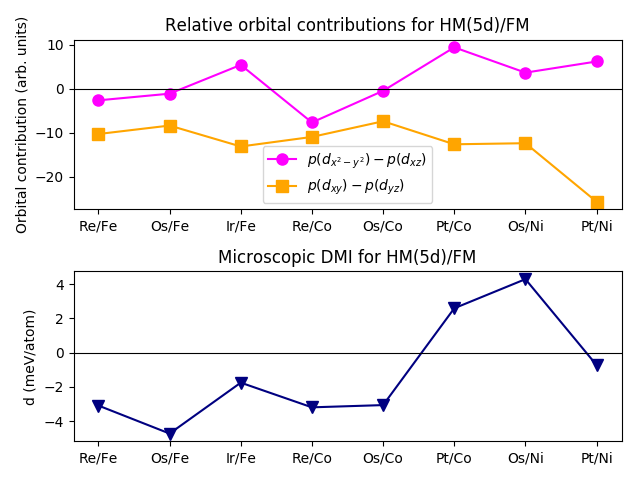}
    \caption{Plot for difference in orbital contributions (around K), and microscopic DMI, for chosen HM(5d)/FM bilayers. The difference between contributions of $d_{x^2-y^2}$ and $d_{xz}$ is plotted with magenta circles, and that between contributions of $d_{xy}$ and $d_{yz}$ is plotted with orange squares. Microscopic DMI is plotted using navy blue triangles for comparision.} 
    \label{fig:orb_contri}
\end{figure}

With this in mind, we compare the contribution of d orbitals to the band structure for some chosen bilayers (Re/Fe, Os/Fe, Ir/Fe, Re/Co, Os/Co, Pt/Co, Os/Ni and Pt/Ni). These bilayers were selected from the bigger data set for their large values of DMI and similar crystal structures. This enabled us to compare them to one another, as well as, easily demonstrate the mechanism behind DMI. We visually inspected the projected bandstructures in these materials, and found that the SOC transitions occur close to the K point in the Brillouin zone (BZ). The specific energy ranges over which the transitions occurred varied for different materials, but were generally found to be between -3 to -1 eV, or -4 to -2 eV below the Fermi level. Fig.\ref{fig:orb_contri} plots the difference in orbital contributions (p-DOS) from HM d orbitals around the K point, in the energy range corresponding to the bilayers. Keeping in mind that orbital contributions can only roughly capture SOC transitions, they turn out to be a good qualitative predictor of the sign of DMI. They also help explain why the three outliers mentioned above have DMI smaller than that predicted by BO alone. 

 With regards to Fig.\ref{fig:orb_contri}, a positive value of $d_1 - d_2$ contribution, signifies a likely transition from $d_1$ to $d_2$ and vice versa. For our chosen bilayers, we infer that $d_{yz} \rightarrow d_{xy}$ and $d_{xz} \rightarrow d_{x^2-y^2}$ transitions will generate negative DMI terms leading to counterclockwise rotation of spin. We observe from Fig.\ref{fig:orb_contri} that all our chosen materials show a $d_{yz} \rightarrow d_{xy}$ transition and a corresponding negative DMI term. However, early HMs tend to show a $d_{xz} \rightarrow d_{x^2-y^2}$ transition (negative DMI term) whereas late HMs show the reverse $d_{x^2-y^2} \rightarrow d_{xz}$ transition (positive DMI term).  The sum of these DMI terms gives the final DMI. In Re/Fe, Os/Fe and Re/Co, both DMI terms are negative leading to a large negative net DMI. However, in Ir/Fe, Pt/Co and Pt/Ni, the two DMI terms have opposite sign and compete with one another. This explains why these outliers have a lower DMI than expected simply from BO. We do not observe any $\langle d_{xz}| \hat{l}_\pm | d_{z^2} \rangle$ transitions. We note that Os/Ni also has competing transitions but shows a larger than expected DMI. Despite this exception, overall, the relative d orbital contributions are a good qualitative predictor of the sign of DMI.

Our work has important implications as to the understanding of DMI and the control of skyrmions. First, our model also explains the correlation between DMI and orbital anisotropy reported recently\cite{Kim_Ono_correlation}. According to our theory, DMI originates in spin mixing orbital transitions caused by SOC. Such transitions along with generating DMI, will also naturally distort orbital shapes and alter orbital moments, leading to the correlation between DMI and orbital asymmetry observed. Second, our model suggests that DMI in bilayers can be controlled by tuning the HM-FM hybridization. A similar concept has been proposed for bulk $\epsilon$Co\cite{Luo_Liu}. Third, the engineering of relative HM-5d orbital fillings, via strain and symmetry breaking, could provide a pathway for controlling the sign of DMI and the chirality of skyrmions. Control of skyrmions with an electric field would be a tremendous leap forward in the development of skyrmion technology.

%-----------CALCULATION METHODS------------
\section{Methods} We calculated the DMI using density functional theory (DFT) for a comprehensive series of HM/FM bilayers. Our choice of HM varied through the 5d and 6p series of elements, ranging from Hf to Bi. We chose the FM to be $\gamma$-Fe(111), Co(0001) or Ni(111). DFT calculations were performed using Vienna Ab-initio Simulation Package (VASP)\cite{VASP1,VASP2,VASP3} with PAW-PBE pseudopotentials\cite{PAW-PBE}. Taking the initial bulk structures from literature, we optimized their volume and constructed bilayers comprising of a monolayer of FM, a monolayer of HM and 10\AA  of vacuum. A schematic of these structures is shown in Fig.\ref{fig:str}. As the DMI in these bilayers is known to be interfacial\cite{Yang_Chshiev_1}, using monolayers of FM and HM is sufficient to capture the essential physics. The in-plane lattice constant was fixed to that of the HM and the FM was strained to less than 5\% to match. The calculations were converged with respect to kmesh and plane wave energy cut offs. All structures were then relaxed till the Hellman-Feynman forces on all atoms were less than 0.01 eV/\AA. DMI was the calculated using the mechanism described by Yang et al.\cite{Yang_Chshiev_2}. The microscopic DMI (\textit{d}) obtained from first-principles was used to further calculate the micromagnetic DMI (\textit{D}). 
An important caveat is that our calculations for DMI are performed with the assumption that the nearest neighbour (NN) DMI term is the dominant contribution to the total DMI. We make this assumption based on the expectation that the hopping terms between NN FM atoms (via HM), will in general be stronger than the hopping terms between next-nearest neighbours (NNN) FM atoms (via HM). According to our model, larger hopping terms will generally lead to stronger DMI. This picture should hold for our structures where there is a significant lattice mismatch between HM and FM layers and strong structural relaxation. It has been shown for chiral bulk materials\cite{Effects_Oliveira}, that NNN DMI terms become less important than dominant NN DMI terms at non zero magnetic fields. Therefore, we expect our results to be relevant to obtaining room temperature, small sized skyrmions at non zero magnetic fields.

\section{acknowledgments}
The authors would like to thank Qian Niu for useful discussions on the microscopic model for DMI and NSF for financial support under the grants NNCI ECCS-1542159, EFRI-newLAW and NASCENT ERC. They also acknowledge the Texas Advanced Computing Center (TACC) at The University of Texas at Austin for providing HPC resources that have contributed to the research results reported within this paper. URL: http://www.tacc.utexas.edu

\section{Author Contributions}
P. J. designed the project, performed the calculations, derived the theory and wrote the initial manuscript. All authors discussed the results and the derivation, as well as contributed to the final version of manuscript.

\section{Figures}

%------------FIGURE 1 -------------

%-----------------FIGURE 2 -----------------

%-----------------FIGURE 3 -----------------

%------------FIGURE 4 -------------

%-----------------FIGURE 5 -----------------


\begin{thebibliography}{39}

\bibitem{Luchaire_Fert} Moreau-Luchaire, C. et al. Additive interfacial chiral interaction in multilayers for stabilization of small individual skyrmions at room temperature. \textit{Nature Nanotechnology} \textbf{11,} 444-448 (2016).

\bibitem{Fert_Cros} Fert, A. Reyren, N. \& Cros, V. Magnetic skyrmions: advances in physics and potential applications. \textit{Nature Reviews Materials} \textbf{2,} 17031 (2017).


\bibitem{Fert_Sampaio} Fert, A. Cros, V. \& Sampaio, J. Skyrmions on the track. \textit{Nature nanotechnology} \textbf{8,} 152 (2013).


\bibitem{Hoffman_Bader} Hoffmann A. \& Bader, S. D. Opportunities at the Frontiers of Spintronics. \textit{Physical Review Applied} \textbf{4,} 047001 (2015).

\bibitem{Bode_Wiesendanger} Bode, M. et al. Chiral magnetic order at surfaces driven by inversion asymmetry. \textit{Nature} \textbf{447,} 190 (2007).


\bibitem{Kang_Zhao} Kang, W. et al. Voltage controlled magnetic skyrmion motion for racetrack memory. \textit{Scientific reports} \textbf{6,} 23164 (2016).


\bibitem{Bogdanov_Yablonskii} Bogdanov, A. N. \& Yablonskii, D. A. Thermodynamically stable ``vortices" in magnetically ordered crystals. The mixed state of magnets. \textit{Zh. Eksp. Teor. Fiz} \textbf{95,} 178-182 (1989).

\bibitem{Pfleiderer} R\"ossler, U. K. Bogdanov, A. N. \& Pfeidereri, C. Spontaneous skyrmion ground states in magnetic metals. \textit{Nature} \textbf{442,} 797 (2006).


\bibitem{Heinze} Heinze, S. et al. Spontaneous atomic-scale magnetic skyrmion lattice in two dimensions. \textit{Nature Physics} \textbf{7,} 713 (2011).


\bibitem{Romming} Romming, N. et al. Writing and deleting single magnetic skyrmions. \textit{Science} \textbf{341,}  636-639 (2013).


\bibitem{Dupe_Heinze} Dup$\acute{e}$, B. , Bihlmayer, G. B\"ottcher, M.  Bl\"ugel, S. \& Heinze, S. Engineering skyrmions in transition metal multilayers for spintronics. \textit{Nature communications} \textbf{7,} 11779 (2016).


\bibitem{Hrabec_McGrouther} Hrabec, A. et al. Measuring and tailoring the Dzyaloshinskii-Moriya interaction in perpendicularly magnetized thin films. \textit{Physical Review B} \textbf{90,} 020402 (2014).


\bibitem{Woo_Beach} Woo, S. et al. Observation of room temperature magnetic skyrmions and their current driven dynamics in ultrathin metallic ferromagnets. \textit{Nature materials} \textbf{15,} 501 (2016).


\bibitem{Soumyanarayanan_Panagopoulos} Soumyanarayanan, A. et al. Tunable room-temperature magnetic skyrmions in Ir/Fe/Co/Pt multilayers. \textit{Nature materials} \textbf{16,} 898 (2017).


\bibitem{Boulle_Gaudin} Boulle, O. et al. Room-temperature chiral magnetic skyrmions in ultrathin magnetic nanostructures. \textit{Nature nanotechnology} \textbf{11,} 449-454 (2016).

\bibitem{Geometrically_Ho} Ho, P. et al. Geometrically Tailored Skyrmions at Zero Magnetic Field in Multilayered Nanostructures. \textit{Phys. Rev. Applied} \textbf{11,} 024064 (2019).


\bibitem{Soumyanarayanan} Soumyanarayanan, A. Reyren, N. Fert, A. \& Panagopoulos, C. Emergent phenomena induced by spin orbit coupling at surfaces and interfaces. \textit{Nature} \textbf{539,} 509 (2016).


\bibitem{Simon_Szunyogh} Simon, E. et al. Spin-correlations and magnetic structure in an Fe monolayer on 5d transition metal surfaces. \textit{Journal of Physics: Condensed Matter} \textbf{26,} 186001 (2014).


\bibitem{Yang_Chshiev_2} Yang, H. Thiaville, A. Rohart, S. Fert, A. \& Chshiev, M. Anatomy of Dzyaloshinskii-Moriya interaction at Co/Pt interfaces. \textit{Physical review letters} \textbf{115,} 267210 (2015).


\bibitem{Yang_Chshiev_1} Yang, H. Boulle, O. Cros, V. Fert, A. \& Chshiev, M. Controlling Dzyaloshinskii-Moriya Interaction via Chirality Dependent Atomic-Layer Stacking, Insulator Capping and Electric Field. \textit{Scientific reports} \textbf{8,} 12356 (2018).


\bibitem{chen2013tailoring} Chen, G. et al. Tailoring the chirality of magnetic domain walls by interface engineering. \textit{Nature communications} \textbf{4,} 2671 (2013).


\bibitem{VASP1} Kresse, G. \& Hafner, J. Ab initio molecular dynamics for liquid metals. \textit{Physical Review B} \textbf{47,} 558 (1993).


\bibitem{VASP2} Kresse, G. \& Furthm\"uller, J. Efficiency of ab-initio total energy calculations for metals and semiconductors using a plane-wave basis set. \textit{Computational materials science} \textbf{6,} 15-50 (1996).


\bibitem{VASP3} Kresse, G. \& Furthm\"uller, J. Efficient iterative schemes for ab initio total-energy calculations using a plane-wave basis set. \textit{Physical review B} \textbf{54,} 11169 (1996).


\bibitem{PAW-PBE} Kresse, G. \& Joubert, D. From ultrasoft pseudopotentials to the projector augmented-wave method. \textit{Physical Review B} \textbf{59,} 1758 (1999).


\bibitem{Reis_Claessen} Reis, F. et al. Bismuthene on a SiC substrate: A candidate for a high-temperature quantum spin Hall material. \textit{Science} \textbf{357,} 287-290 (2017).


\bibitem{Tolley_Fullerton} Tolley, R. Montoya, S. A. \& Fullerton, E. E. Room-temperature observation and current control of skyrmions in Pt/Co/Os/Pt thin films \textit{Physical Review Materials} \textbf{2,} 044404 (2018).


\bibitem{Dzyaloshinsky} Dzyaloshinsky, I. A thermodynamic theory of ``weak" ferromagnetism of antiferromagnetics. \textit{Journal of Physics and Chemistry of Solids} \textbf{4,}  241-255 (1958).


\bibitem{Moriya} Moriya, T.  Anisotropic superexchange interaction and weak ferromagnetism. \textit{Physical Review} \textbf{120,} 91-98 (1960).


\bibitem{Smith} Smith, D. A. New mechanisms for magnetic anisotropy in localised S-state moment materials. \textit{Journal of Magnetism and Magnetic Materials} \textbf{1,} 214-225 (1976).


\bibitem{Fert_Levy1} Fert, A. \& Levy, P. M. Role of anisotropic exchange interactions in determining the properties of spin-glasses. \textit{Physical Review Letters} \textbf{44,} 1538 (1980).


\bibitem{Fert_Levy2} Fert, A. R. Magnetic and transport properties of metallic multilayers. \textit{Materials Science Forum.} \textbf{59-60} 439-480 (1990).


\bibitem{Heide_Blugel} Heide, M.  Bihlmayer, G. Mavropoulos, Ph. Bringer, A. \& Bl\"ugel, S. Spin Orbit Driven Physics at Surfaces. \textit{Newsletter of the Psi-K Network} \textbf{78,} 1 (2006).

\bibitem{Ryu} Ryu, K.-S. Yang, S.-H., Thomas, L. \& Parkin, S. S. P. Chiral spin torque arising from proximity-induced magnetization. \textit{Nature communications} \textbf{5,} 3910 (2014).


\bibitem{Belabbes_Manchon} Belabbes, A. Bihlmayer, G. Bechstedt, F.  Bl\"ugel, S. \& Manchon, A. Hund's Rule-Driven Dzyaloshinskii-Moriya Interaction at 3d-5d Interfaces. \textit{Physical review letters} \textbf{117,} 247202 (2016).


\bibitem{Freimuth} Freimuth, F. Bl\"ugel, S. \& Mokrousov, Y. Berry phase theory of Dzyaloshinskii-Moriya interaction and spin-orbit torques. \textit{Journal of physics:Condensed matter} \textbf{26,} 104202 (2014).

\bibitem{Luo_Liu} Luo, H.-B. Zhang, H.-B. \& Liu, J. P. Strong hopping induced Dzyaloshinskii–Moriya interaction and skyrmions in elemental cobalt \textit{NPJ Computational Materials} \textbf{5,} 50 (2019). 


\bibitem{Hanke_Mokrousov} Hanke, J.-P. Freimuth, F. Bl\"ugel, S. \& Mokrousov, Y. Higher-dimensional Wannier Interpolation for the Modern Theory of the Dzyaloshinskii-Moriya Interaction: Application to Co-based Trilayers. \textit{Journal of the Physical Society of Japan} \textbf{87} 041010 (2018).

\bibitem{Jiang_Hoffmann} Jiang, W. et al. Skyrmions in magnetic multilayers \textit{Physics Reports} \textbf{704,} 1-49 (2017).

\bibitem{Kim_Ono_correlation} Kim, S. et al. Correlation of the Dzyaloshinskii–Moriya interaction with Heisenberg exchange and orbital asphericity. \textit{Nat. Comm.} \textbf{9,} 1648 (2018).

\bibitem{frust_th1} Okubo, T. Chung, S. and Kawamura, H. Multiple-q States and the Skyrmion Lattice of the Triangular-Lattice Heisenberg
Antiferromagnet under Magnetic Fields. \textit{Phys. Rev. Lett.} \textbf{108,} 017206 (2012).

\bibitem{frust_th2} Leonov, A.O and Mostovoy, M. Multiply periodic states and isolated skyrmions in an anisotropic frustrated magnet. \textit{Nat. Comm.} \textbf{6,} 8275 (2015).

\bibitem{frust_th3} Hayami, S. Lin, S.-Z. and Batista, C. D. Bubble and skyrmion crystals in frustrated magnets with easy-axis anisotropy. \textit{Phys. Rev. B} \textbf{93,} 184413 (2016).

\bibitem{frust_th4} Liang, J. J. et al. Magnetic field gradient driven dynamics of isolated skyrmions and antiskyrmions in frustrated magnets. \textit{New J. Phys.} \textbf{20,} 053037 (2018).

\bibitem{frust_ex1} Kurumaji, T. et al. Skyrmion lattice with a giant topological Hall effect in a frustrated triangular-lattice magnet. \textit{Science} \textbf{10,} 1126 (2019).

\bibitem{Engineering_Dupe} Dup$\acute{e}$, B. Bihlmayer, G. B$\ddot{o}$ttcher, M. Bl$\ddot{u}$gel, S. and Heinze, S. Engineering skyrmions in transition-metal multilayers for spintronics. \textit{Nat. Comm.} \textbf{7,} 11779 (2016).

\bibitem{Thickness_Cho} Cho, J. et al. Thickness dependence of the interfacial Dzyaloshinskii–Moriya interaction in inversion symmetry broken systems. \textit{Nat. Comm.} \textbf{6,} 7635 (2015).

\bibitem{Kashid_Salunke} Kashid, V. Shah, V. and Salunke, H. G. Dzyaloshinskii-Moriya interaction and chiral magnetism in 3d-5d zigzag chains: Tight-binding model and ab initio calculations. \textit{Phys. Rev. B} \textbf{90,} 054412 (2014).

\bibitem{Effects_Oliveira} Oliveira, E. A. S. Silva, R. L. Silva, R. C. and Pereira, A. R. Effects of second neighbor interactions on skyrmion lattices in chiral magnets. \textit{J. Phys.: Condens. Matter} \textbf{29,} 205801 (2017).

\bibitem{Interfacial_Belmeguenai} Belmeguenai, M. et al. Interfacial Dzyaloshinskii-Moriya interaction in perpendicularly magnetized $Pt/Co/AlO_x$ ultrathin films measured by Brillouin light spectroscopy. \textit{Phys. Rev. B} \textbf{91,} 180405 (2015).


\end{thebibliography}
\end{document}